\input{aipcheck}

\documentclass[
    ,final            
  ]
  {aipproc}

\layoutstyle{6x9}

\begin{document}
\newcommand{\be}{\begin{equation}}
\newcommand{\ee}{\end{equation}}
\newcommand{\bea}{\begin{eqnarray}}
\newcommand{\eea}{\end{eqnarray}}
\newcommand{\bT}{{${\bf b}_\perp$}}
\newcommand{\xT}{{${\bf x}_\perp$}}

\title{GPDs and DVCS with Positrons}


\author{Matthias Burkardt}{
  address={Department of Physics,
New Mexico State University, Las Cruces, NM 88003, and},altaddress={
Jefferson Lab, 12000 Jefferson Ave., Newport News, VA 23606.}
}

\begin{abstract}
The beam charge asymmetry helps to isolate the real part of the
deeply virtual Compton scattering (DVCS) amplitude. 
It is discussed what information can
be gained both from the real and imaginary part of the DVCS amplitude.
\end{abstract}

\maketitle

\section{Introduction}
\begin{figure}
\resizebox{.8\textwidth}{!}{\includegraphics{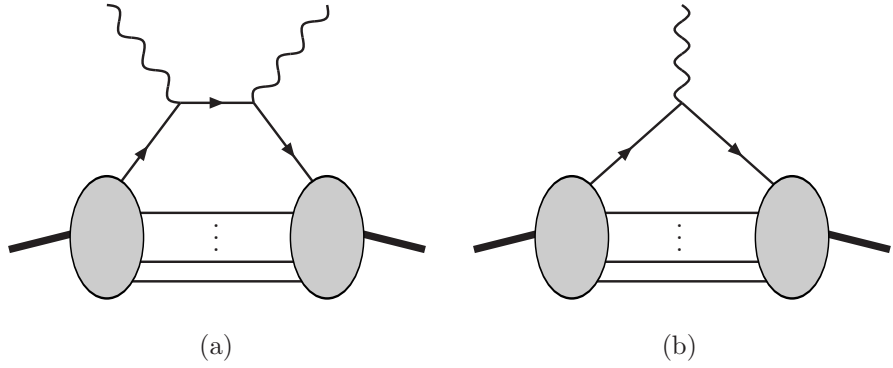}}
\caption{Comparison between Compton scattering (a)
and form factors (b).
The Compton amplitude (handbag diagrams dominate for
large virtualities of the virtual photon), involves a quark propagator
between the two photon vertices}
\label{fig:DVCS}
\end{figure}

Form factors are the coherent amplitude that the nucleon remains
intact when one of its quarks absorbs a certain momentum transfer
$\Delta^\mu$. In this amplitude, 
the contribution from all quarks carrying
all kinds of momenta is added up coherently. 
Compton scattering provides a
more surgical approach (Fig. \ref{fig:DVCS}). 
Due to the presence of the quark propagator 
between the two photon vertices (note that for virtual photons with
large virtuality, the Compton amplitude is dominated by handbag 
diagrams), the Compton amplitude is sensitive
to the momentum fraction $x$ carried by the active quark.

Suppose one could disect the Dirac form factor $F_1^q$ or the
Pauli form factor $F_2^q$ for quarks with flavor $q$ w.r.t. the 
(average) momentum fraction
$x=\frac{1}{2}\left(x_i+x_f\right)$ carried by the active quark.
The result would be the generalized parton distributions
$H^q(x,\xi,t)$ and $E^q(x,\xi,t)$ respectively
\be
F_1^q(t)=\int_{-1}^1 dx H^q(x,\xi,t) \quad \quad \quad
F_2^q(t)=\int_{-1}^1 dx E^q(x,\xi,t),
\label{eq:disect}
\ee
where $t=\Delta^2$.
As the variable $x$ represents the (average)
momentum fraction of the active quark in the
`infinite momentum' direction, it makes a difference whether
the momentum transfer is parallel or perpendicular to that direction.
This `angular dependence' is provided by the dependence of GPDs
on the variable $\xi = \frac{1}{2}\left(x_f-x_i\right)$.
Since the information about the infinite momentum direction is
`lost' when the quark momentum is not specified,
the $\xi$-dependence of GPDs must disappear when $x$ is integrated 
over, as in (\ref{eq:disect}). Also a consequence of Lorentz 
invariance, a whole set of `polynomiality conditions' exists
for higher moments of GPDs. For example, the $x^{n-1}$-moments of
GPDs, with $n$ even, are even polynomials in $\xi$ with highest
power equal to $n$
\be
\int_{-1}^1 dx\, x^{n-1} H^q(x,\xi,t) = A^q_{n0}(t) + A^q_{n2}(t)\xi^2
+...+ A^q_{nn}(t)\xi^n,
\ee
and similar for $E^q(x,\xi,t)$. These conditions provide
highly nontrivial constraints for the $x,\xi$-dependence and may
play a crucial role in the GPD extraction from the DVCS amplitude.

\begin{figure}
\resizebox{.8\textwidth}{!}{\includegraphics{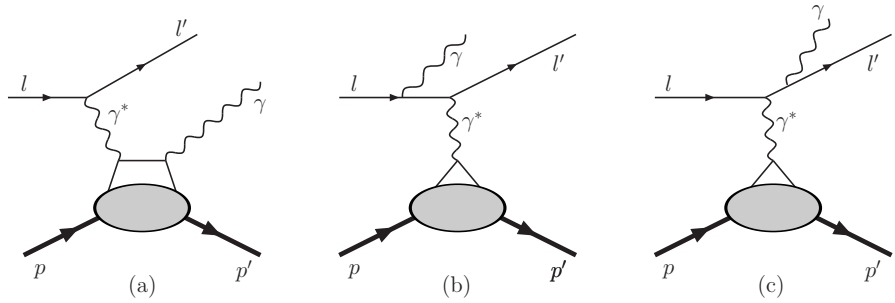}}
\label{fig:BH}
\caption{Interference between the DVCS amplitude (a) and the 
Bethe-Heitler process (b,c).}
\end{figure}

GPDs enter the DVCS amplitude ${\cal A}_{DVCS}$ through convolution integrals
\be
{\cal A}_{DVCS} \sim \int_{-1}^1 dx 
\frac{GPD(x,\xi,t)}{x-\xi+i\varepsilon},
\label{eq:conv}
\ee
i.e. the imaginary part of the DVCS amplitude depends only on
GPDs along the `diagonal' $x=\xi$
\be
\Im {\cal A}_{DVCS}(\xi,t) \sim GPD(\xi,\xi,t),
\label{eq:Im}
\ee
while the real part also probes GPDs for $x\neq \xi$
\be
\Re {\cal A}_{DVCS}(\xi,t) \sim \int_{-1}^1 dx 
\frac{GPD(x,\xi,t)}{x-\xi}.
\label{eq:Re}
\ee
Experimentally, the DVCS amplitude interferes with the
Bethe-Heitler process (Fig.\ref{fig:BH})
and 
\be
\sigma = \left|{\cal A}_{BH}+{\cal A}_{DVCS}\right|^2
=\left|{\cal A}_{BH}\right|^2
+\left|{\cal A}_{DVCS}\right|^2
+ 2\Re \left\{{\cal A}_{BH}{\cal A}_{DVCS}^*\right\}
\ee
while $\Im {\cal A}_{DVCS}(\xi,t)$ can be separated using the beam 
spin
asymmetry, and $\Re {\cal A}_{DVCS}(\xi,t)$ from the angular 
dependence, the {beam charge asymmetry} ($e^+$ v. $e^-$ or
$\mu^+$ v. $\mu^-$) can be used as an independent means to
isolate the real part of the
DVCS amplitude:
the interference terms between BH and
DVCS depends on the sign of the lepton charge!
For more details regarding the measurement of the DVCS amplitude,
see Ref. \cite{RPP} and references therein.

\section{Physics of GPDs}
One of the reasons  GPDs attracted interest is that
the $2^{nd}$ moment of $H^q+E^q$ can be identified with
the angular momentum (spin plus orbital)
carried by quarks with flavor $q$
in a nucleon with spin polarization ${\vec S}$ \cite{Ji}
\be
\langle {\vec J}^q \rangle 
\equiv \langle q^\dagger {\vec \Sigma}q\rangle
+\langle q^\dagger {\vec r}\times\left({\vec p}-q{\vec A}\right)
q\rangle = {\vec S} \int_{-1}^1 dx\,x\left[H^q(x,\xi,0)+E^q(x,\xi,0)
\right].
\label{eq:Ji}
\ee
Note that while the $\xi$-dependence (term $\propto \xi^2$)
on the r.h.s. cancels between $H^q(x,\xi,0)$ and $E^q(x,\xi,0)$,
those integrals must be performed at the same fixed value of $\xi$.

Another important aspect of GPDs is their connection with
{\em impact parameter dependent parton distributions} 
$q(x,{\bf b}_\perp)$. The latter are defined similar to the usual 
PDFs, but at distance ${\bf b}_\perp$ relative to the
{\em transverse center of longitudinal momentum} ${\bf R}_\perp
\equiv \sum_{i\in q,g}x_i{\bf b}_{\perp,i}$. The Fourier transform of
GPDs for $\xi=0$ yields these impact parameter dependent PDFs. 
For example, the two-dimensional Fourier transform of 
$H^q(x,0,-{\Delta}_\perp^2)$ yields the distribution of
unpolarized quarks in an unpolarized or longitudinally polarized
hadron \cite{mb1}
\be
q(x,{\bf b}_\perp) = \int \frac{d^2 {\Delta}_\perp}{(2\pi)^2}
H^q(x,0,-{\Delta}_\perp^2) e^{-i{\bf b}_\perp \cdot 
{\Delta}_\perp}.
\ee
The transverse gradients of the Fourier transform of 
$E^q(x,0,-{\Delta}_\perp^2)$ describes the transverse deformation
of the distribution of
unpolarized quarks in a transversely polarized nucleon \cite{IJMPA}.
For more details, see Ref. \cite{RPP} and references 
therein.

Unfortunately, the DVCS amplitude is mostly sensitive to the
regime $x\approx \xi$. In fact, the imaginary part is sensitive
to $x=\xi$ only (\ref{eq:Im}), while the convolution integral for the
real part (\ref{eq:Re}) is dominated by the vicinity of
$x\approx \xi$.

\section{Information Content of the DVCS Amplitude}
In the following we will focus on charge even GPDs 
$GPD^{+}(x,\xi,t)\equiv GPD(x,\xi,t)-GPD(-x,\xi,t)$. Furthermore,
the $Q^2$ dependence will not be shown explicitly although all
terms depend on $Q^2$.

Using dispersion relations one can show that \cite{DI}
\be
\Re {\cal A}_{DVCS} \sim \int_{-1}^1 dx \frac{GPD^+(x,\xi,t)}{x-\xi}
= \int_{-1}^1 dx \frac{GPD^+(x,x,t)}{x-\xi} + \Delta(t),
\label{eq:DR}
\ee
where $\Delta(t)$ is the D-form factor \cite{PW}. This remarkable 
relation
also follows from polynomiality \cite{AT} and implies that the
information content of the DVCS amplitude (at fixed $Q^2$) can be
condensed to GPDs along the diagonal $x=\xi$ plus the D-form factor.
It should be emphasized that $\Re {\cal A}_{DVCS}(\xi,t)$ still adds
more information to $\Im{\cal A}_{DVCS}(\xi,t)$ than just $\Delta(t)$,
since, for fixed $t$ not the whole range $0<\xi<1$ is  
accessible (at very low $\xi$ the Bjorken limit may not yet have been
reached  and high $\xi$ is inaccessible since
$t=-\frac{4\xi^2M^2+\Delta_\perp^2}{1-\xi^2}$), while the above 
integrals extend from $\xi=0$ to $\xi=1$. Nevertheless, (\ref{eq:DR})
suggests to fit parameterizations for $GPD(\xi,\xi,t)$ and $D(t)$
to DVCS data rather than attempting to constrain parameters for
$GPD(x,\xi,t)$ over the whole $x-\xi$ range. $GPD(\xi,\xi,t)$ and 
$\Delta(t)$ could then be used as an interface between experimental 
data
and models. In fact, due to (\ref{eq:DR}), any model/parameterization
of GPDs satisfying polynomiality that fits both $\Re {\cal A}_{DVCS}$
and  $\Im {\cal A}_{DVCS}$, would also fit $GPD(\xi,\xi,t)$ and 
$\Delta(t)$ and vice versa. Moreover, fitting GPD-models to
DVCS-data is not unique. For example, one can always 
fit DVCS data with the ansatz
\be
H(x,\xi,t)=H_{DD}(x,\xi,t)+ 
\Theta(|\xi|-|x|)D\left(\frac{x}{\xi},t\right)
\label{eq:DD}
\ee 
where for the `double distribution' \cite{DD} one makes
the specific ansatz $H_{DD}(x,\xi,t)=H(x,x,t)$, and $D(z,t)$ 
being an arbitrary function that satisfies 
$\int_{-1}^1 \frac{dz}{z-1}D(z,t)=\Delta(t)$.

With the information from DVCS reduced to GPDs along the diagonal
plus the $D$-form factor, the question arises what one can learn
from this information. GPDs along the diagonal have been discussed
e.g. in Ref. \cite{mb:xi}. 

One very interesting observable that one can extract from this 
information is the $\frac{1}{x}$-moment of GPDs for $\xi=0$.
Provided $t$ is large enough that the limit exists, one can take the
$\xi\rightarrow 0$ limit in the remarkable relation on the r.h.s.
of (\ref{eq:DR}), yielding
\be
\int_{-1}^1 dx \frac{GPD^+(x,0,t)}{x}
= \int_{-1}^1 dx \frac{GPD^+(x,x,t)}{x} + \Delta(t).
\label{eq:WACS}
\ee
From the GPD along the diagonal plus the $D$-form factor one can thus
obtain the same $\frac{1}{x}$-moment of GPDs with  $\xi=0$ that also
enters the wide angle Compton scattering (WACS) \cite{WACS}
amplitudes. The main advantage of the approach
outlined here compared to WACS is that using DVCS and 
(\ref{eq:WACS}), one can
access $\int_{-1}^1 dx \frac{GPD^+(x,0,t)}{x}$ in a regime where
$M^2<-t\ll Q^2$, where the clear seperation of scales facilitates
the interpretation in terms of factorization.

Knowledge of $\int_{-1}^1 dx \frac{GPD^+(x,0,t)}{x}$ would be
very valuable for understanding the physics of form factors at 
large $-t$: since antiquarks and sea quarks are not expected
to play a significant role at large $-t$, a flavor seperation
of the $u$ and $d$ contributions should be possible using proton 
and neutron data only. Comparing 
$\int_{-1}^1 dx \frac{GPD^+(x,0,t)}{x}$ for each quark flavor
with flavor seperated form factor date should help understand
what type of quarks dominate those form factors at large momentum 
transfer. For example, if the ratio between 
$\int_{-1}^1 dx \frac{GPD^+(x,0,t)}{x}$
and the corresponding form factor approaches 1 at large $-t$ then
that form factor would be dominated by quarks at $x\rightarrow 1$
in that limit, while if that ratio turns out to be significantly
greater than 1 for large $-t$
then the corresponding form factor would be dominated by quarks
carrying intermediate $x$.
 
The interpretation of the $D$-form factor
remains obscure. While an interpretation of the $z$-moment
of the $D$-term has been provided in terms of the
stress-tensor \cite{PW}, the $D$-form factor is the $\frac{1}{z}$
moment of the $D$-term. This $\frac{1}{z}$ moment also contributes
a $\delta(x)$-type contribution to $\Re {\cal A}_{DVCS}$ in the
$\xi\rightarrow 0$ limit, since 
$H(x,\xi,t)=H_{DD}(x,\xi,t)+ 
\Theta(|\xi|-|x|)D\left(\frac{x}{\xi}\right)$ on the one hand,
i.e. for $x\neq 0$ one finds $\lim_{\xi\rightarrow 0} 
H(x,\xi,t)=H_{DD}(x,\xi,t)$, but on the other hand
$\lim_{\xi\rightarrow 0}\Re {\cal A}_{DVCS}(\xi,t) =
\int_{-1}^{1}\frac{dx}{x}H(x,0,t) = 
\int_{-1}^{1}\frac{dx}{x}H_{DD}(x,0,t)=\Delta(t)$, i.e.
\be
\frac{H(x,0,t)-H_{DD}(x,0,t)}{x} = \delta(x)\Delta(t).
\ee
\section{$Q^2$-evolution}
One possibility to resolve the non-uniqueness of GPD extraction
from DVCS data is the study of `double DVCS' (DDVCS), where the
photon on the `final state' is also virtual, i.e. rather than
producing a real photon, for example a lepton pair is produced.
An alternative to this difficult process might be using
the $Q^2$ evolution of GPDs. To illustrate this point, imagine
in the study of ordinary PDFs one were able to measure PDFs only at 
one value of $x$, but over a wide range of $Q^2$. Since 
the DGLAP evolution equations that govern the $Q^2$ dependence
of PDFs are known, one could thus still (partly) reconstruct
the PDFs over a broad $x$ range from this information.

In the context of GPDs, one also knows the evolution equations.
Therefore, even if one can access GPDs only along the line 
$x=\xi$, since the $x$-distribution changes under $Q^2$, and since
the Kernels that govern this evolution are also known,
one can use the $Q^2$ dependence to help disentangle the 
$x$-dependence.

Of course, for this procedure to work, even the lowest values 
of $Q^2$ used in such a fit would have to be high enough
to ensure that higher twist effects --- which are usually not 
accounted for in evolution equations --- are certain to be absent.

\begin{theacknowledgments}
This work was supported by the DOE under grant numbers 
DE-FG03-95ER40965 and DE-AC05-06OR23177 (under which Jefferson
Science Associates, LLC, operates Jefferson Lab).
\end{theacknowledgments}


\bibliographystyle{ws-procs9x6}

\end{document}